\documentclass{jfm}

\usepackage{graphicx}
\usepackage{newtxtext}
\usepackage{newtxmath}
\usepackage{natbib}
\usepackage{hyperref}
\hypersetup{
    colorlinks = true,
    urlcolor   = blue,
    citecolor  = black,
}

\usepackage{bm}
\usepackage{amsmath}
\usepackage{nicefrac}       
\usepackage{microtype}      
\usepackage[font=small, labelfont=bf]{subcaption}
\usepackage{adjustbox}
\usepackage{tikz}
\usetikzlibrary{quotes, arrows.meta, calc}
\usepackage{pgfplots}
\usepackage[toc]{appendix}

\begin{document}

\title{Concerning the Use of Turbulent Flow Data for Machine Learning}

\author{M. Sardar\aff{1}
 \corresp{\email{mohammed.sardar@manchester.ac.uk}},
  M. J. Zimo\'n \aff{2}\aff{3},
  S. Draycott\aff{1},
  A. Revell
  \and A. Skillen}

\affiliation{
  \aff{1}School of Engineering, The University of Manchester \\
  \aff{2} IBM Research Europe, Daresbury, UK \\
  \aff{3} School of Mathematics, The University of Manchester 
 }

\maketitle

\begin{abstract}
This article describes some common issues encountered in the use of Direct Numerical Simulation (DNS) turbulent flow data for machine learning. We focus on two specific issues; 1) the requirements for a fair validation set, and 2) the pitfalls in downsampling DNS data before training. We attempt to shed light on the impact these issues can have on machine learning and computer vision for turbulence. Further, we include statistical and spectral analysis for the homogenous isotropic turbulence from the John Hopkins Turbulence Database, a Kolmogorov flow, and a Rayleigh-B\'enard Convection Cell using data generated by the authors, to concretely demonstrate these issues. 
\end{abstract}

\section{Introduction}

Throughout the last 10 years, the fluid dynamics research community has increased its focus on Machine Learning (ML). The broad application of Deep Learning (DL) and Convolutional Neural Network (CNN) based computer vision techniques to the study of turbulent flows \citep{pandey2020, yu2023, calzolari2021, saezdeocarizborde2022, sandberg2022} has brought with it a new set of challenges for the turbulence community. 

Here, we focus on key issues which can arise in the application of common ML methodologies to Direct Numerical Simulation (DNS) outputs. We consider the impact these methodologies may have on accuracy, and generalisability, and how this can affect results from ML work. Importantly, we present circumstances where results may appear to demonstrate superior performance against state of the art results, but are biased. 


\section{On Correlated-in-time Datasets}
\subsection{In Prior Work}
\label{subsec: corr prior work}

ML approaches for computer vision tasks such as classification, segmentation, and super-resolution typically assume that samples of data in a dataset are Independent and Identically Distributed (IID). CNNs, which form the backbone of of most modern computer vision tasks \citep{zhao2024}, share this assumption. However, in practice, this assumption is often violated, especially in real-world data. Despite this, ML models, particularly deep learning models, have demonstrated remarkable success, suggesting that the IID assumption is a soft constraint \citep{xiong2019}.

\begin{figure}
\centering
\includegraphics[width=0.4\textwidth]{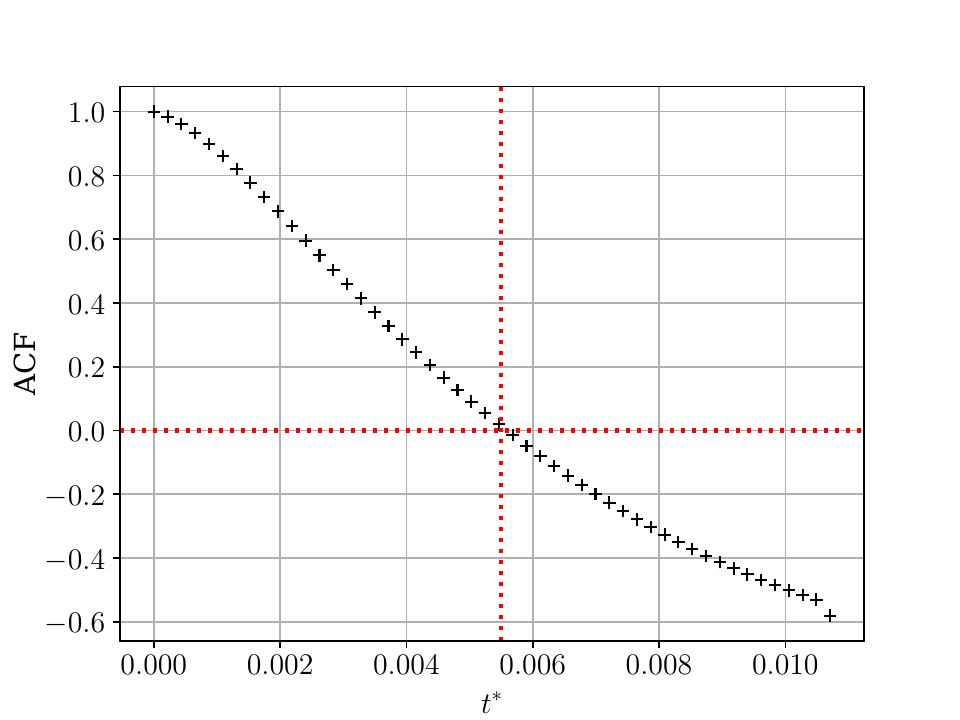}
\setlength{\belowcaptionskip}{-10pt}
\caption{An analysis of the autocorrelation function for the JHTDB isotropic data, over 100 snapshots in time, separated by $10\Delta t$, where $\Delta t=0.0002$ is the DNS timestep.}
\label{fig:jhtdb acf}
\end{figure}

The widespread use of CNNs in tasks related to fluid dynamics \citep{pandey2020} illustrates that the same condition is violable in applications to turbulence. Not only does turbulence not adhere to the independence assumption, but turbulent structures may be correlated to one another spatially (measured by a two-point correlation), and correlations in time may exist between individual snapshots of a given DNS instance for any turbulent flow \citep{pope2000}. We define independence for two random variables $X$ and $Y$ to be when their joint probability distribution, $p_{X, Y}$,  is the product of their marginal distributions, i.e., $p_{X, Y}\left(X, Y\right) = p_X\left(x\right)p_Y\left(y\right)$ for all $x$ and $y$. The decorrelation is defined as the correlation coefficient being zero; this leads to the
expectation of the product of two independent random variables being equal to the product of individual expectations, i.e., $\mathbf{E}\left[XY\right] = \mathbf{E}[X]\mathbf{E}[Y]$. We focus here on correlations in time, defined via the Autocorrelation Function (ACF) Eq. \ref{eq: acf}, valid for statistically stationary flows. For statistically non-stationary flows, transformations exist which may be applied to the data to extract statistically stationary data (e.g. differencing). The ACF is defined here as:

\begin{equation}
f_{ac} \left( s \right) = \frac{\left<u(t)u(t+s)\right>}{\left<u(t)^2 \right>},
\label{eq: acf}
\end{equation}
where $f_{ac}$ is the autocorrelation function, $u(t)$ is the velocity at a time $t$, $s$ is the time-lag, and $\left<\cdot\right>$ represents an averaging operator. Figure \ref{fig:jhtdb acf} provides an illustrative ACF for the homogeneous isotropic turbulence dataset from the John Hopkins Turbulence Database (JHTDB), where $t^*$ is the timestep non-dimensionalised as $t^* = \nicefrac{tU}{L}$, in which $U=0.686$ is the RMS velocity, and $L=2\pi$ is the domain length \citep{perlman2007, yeung2012}. 

In prior work, flow-specific physical time scales are often used as a means of ensuring samples are temporally decorrelated. The separation period may be defined in terms of viscous time scales \citep{jagodinski2023}, wall time units (for wall-bounded flows) \citep{nakamura2021}, integral time scales \citep{kim2020a} and large-eddy turnover time ($t_L$) \citep{kim2021a}. 
In other cases, separation time is expressed in terms of simulation time-step, such as $10\Delta t_{DNS}$ \citep{yousif2022}, $100\Delta t_{DNS}$ \citep{hasegawa2020},  $200\Delta t_{DNS}$ \citep{guan2022}, and $1000 \Delta t_{DNS}$ \citep{srinivasan2024}. A discussion of the separation time between snapshots is omitted altogether in other pieces of work \citep{fukami2019b,  razizadeh2020, saezdeocarizborde2022}.

In those cases where either the snapshot separation period is based on the numerical (DNS) timestep without an investigation into the separation required for decorrelation of turbulent fluctuations, or a discussion on the separation period is omitted entirely, two considerations emerge:


\begin{enumerate}
  \item Selecting training and validation data as random subsets of the data, snapshots may be highly correlated with one another (i.e. $\Delta t$ between snapshots does not lead to the ACF decaying to zero). This is akin to validating on the training data -- the validation data is not useful in gauging whether a model has overfit to the training data \citep{10.5555/1162264}.
  \item Where correlations between training and validation data are mitigated, correlations within the training set may reduce the `effective' dataset size. Across training samples, the model is not presented with new information to learn from, and may overfit. \citet{kim2021a, subramaniam2020a}, demonstrate that small, decorrelated datasets are sufficient to train benchmark turbulence super-resolution models. We advocate the use of validation data derived from different initial conditions, e.g. where the simulation begins with perturbed initial conditions, for simulations where the initial conditions are a random field.
\end{enumerate}

Physical time scales such as $t_L$ may be large, presupposing that the largest eddies in a flow correspond to the characteristic length scale. In certain cases, such as Kolmogorov flows, prescribed forcing frequencies do not allow for the presence of eddies at those scales. We suggest a heuristic decorrelation metric: once could filter simulation data by computing the ACF and determining the period required for the ACF to decay to zero, obtaining decorrelated snapshots.

Intuitively, statistical correlation differs from dependence. Two snapshots of turbulence with a large $\Delta t$ between them may be completely decorrelated, but both both of them belong to the same time-evolution, and thus the later snapshot is necessarily dependent on the earlier state. We hypothesise that mitigating temporal correlation would improve model performance and training convergence.

\subsection{Results from a DDPM-based Super-Resolution of a Kolmogorov Flow}
We illustrate the impact of training with correlated-in-time data on a case study: Super-resolving snapshots of a 2D Kolmogorov flow using a Denoising Diffusion Probabilistic Model (DDPM) \citep{hoDenoisingDiffusionProbabilistic2020, ho2022}. As the efficacy of DDPM-based generative modelling for turbulence is an active area of research \citep{guo2024, sardar2024, li2024c}, we use this method to demonstrate this issue. A Kolmogorov flow is a theoretical forced isotropic flow, governed by the incompressible Navier-Stokes equations with an additional forcing term:
\begin{flalign}
&\frac{\partial u_i}{\partial t} + u_j\frac{\partial u_i}{\partial x_j} = -\frac{\partial p}{\partial x_i} + \frac{1}{\textrm{Re}}\frac{\partial^2 u_i}{\partial x_j \partial x_j} + f_i, && \\
& \frac{\partial u_i}{\partial x_i} = 0,
\end{flalign}
where $u_i$ is the velocity, $x_i$ represents spatial coordinates, Re is the Reynolds number (set as $\textrm{Re}=222$ in this simulation), $p$ is the pressure, $f_i$ is a sinusoidal forcing term in the $x_2$ direction, and all fields are dimensionless. The governing equations are solved using a spectral code \citep{dresdner2023}, with fully periodic boundaries over a square of length $2\pi$, discretised into $256\times256$ grid points.  

Two datasets are generated: the first, with snapshots separated by the decorrelation period, and the second, with correlated-in-time snapshots, where every $8\Delta t$ corresponds to one decorrelation period. Each dataset is used to train a DDPM for an $8 \times$ super-resolution task, and evaluated on decorrelated data generated using different initialisations to those used in training. Identical architectures and training hyperparameters are used for both cases. We denote low-resolution with LR, super-resolution outputs with SR, and DNS data with HR throughout.

\begin{figure}
\centering\captionsetup{width=\linewidth}
\begin{subfigure}{0.8\columnwidth}
\includegraphics[width=\textwidth]{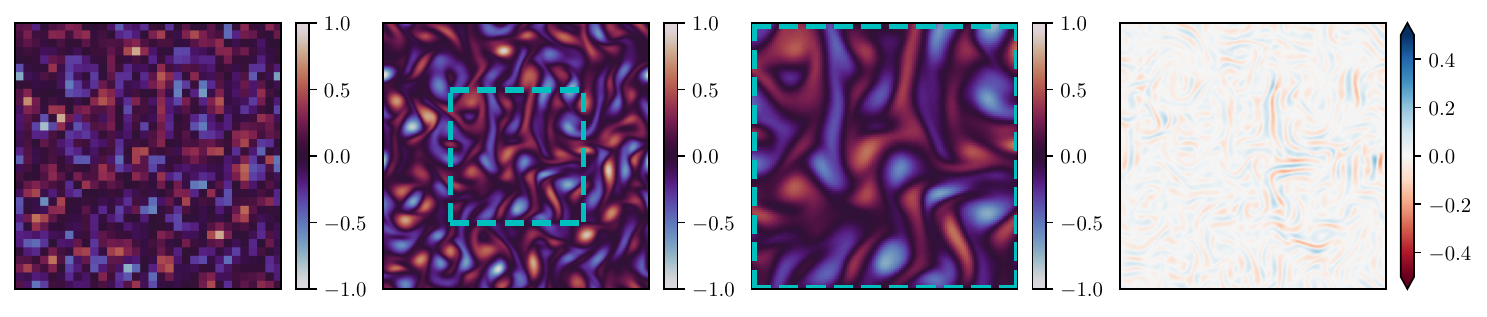}
  \caption{Decorrelated Data}
\end{subfigure}

\begin{subfigure}{0.8\columnwidth}
\includegraphics[width=\textwidth]{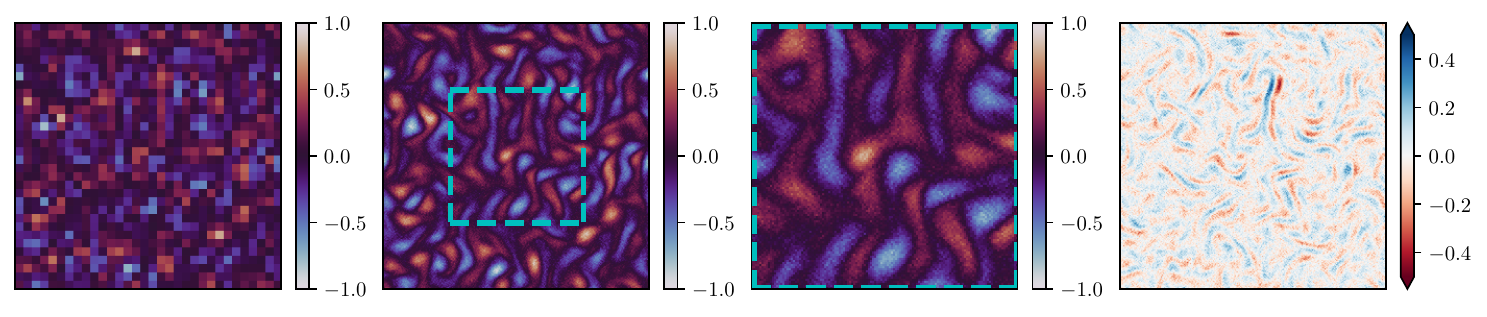}
  \caption{Correlated-in-time Data}
\end{subfigure}
\setlength{\belowcaptionskip}{-10pt}
\caption{Instantaneous snapshots of vorticity, showing the $32\times32$ low-resolution (LR, first), $256\times256$ super-resolved (SR, second), $128\times128$ subcrop of super-resolved field (third), and $256\times256$ $(\textrm{HR} - \textrm{SR})$ (fourth).}
\label{fig: kol snapshots}
\end{figure}

We observe that the model trained on correlated-in-time data suffers when testing on unseen data. This is most evident in the instantaneous snapshots (Figure \ref{fig: kol snapshots}), where it can be noted that the fields generated using the model trained on correlated data contain significantly more noise than the same field generated using the model trained on decorrelated data.

\begin{figure}
\centering\captionsetup{width=\linewidth}
\begin{subfigure}{0.49\textwidth}
\includegraphics[width=\textwidth, trim=5 5 5 5, clip]{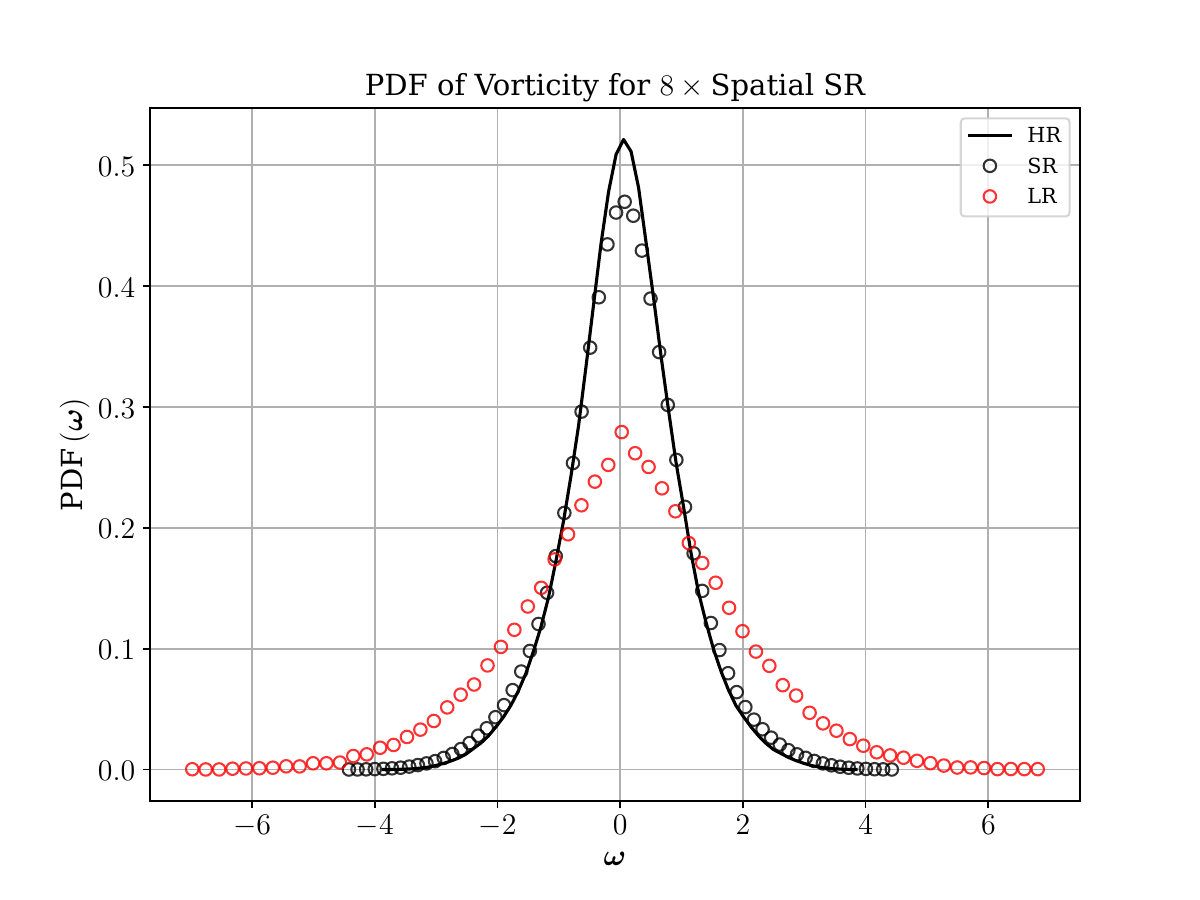}
  \setlength{\belowcaptionskip}{-3pt}
  \caption{Decorrelated Data}
  \label{fig: kol pdf good}
\end{subfigure}
\begin{subfigure}{0.49\textwidth}
\includegraphics[width=\textwidth, trim=5 5 5 5, clip]{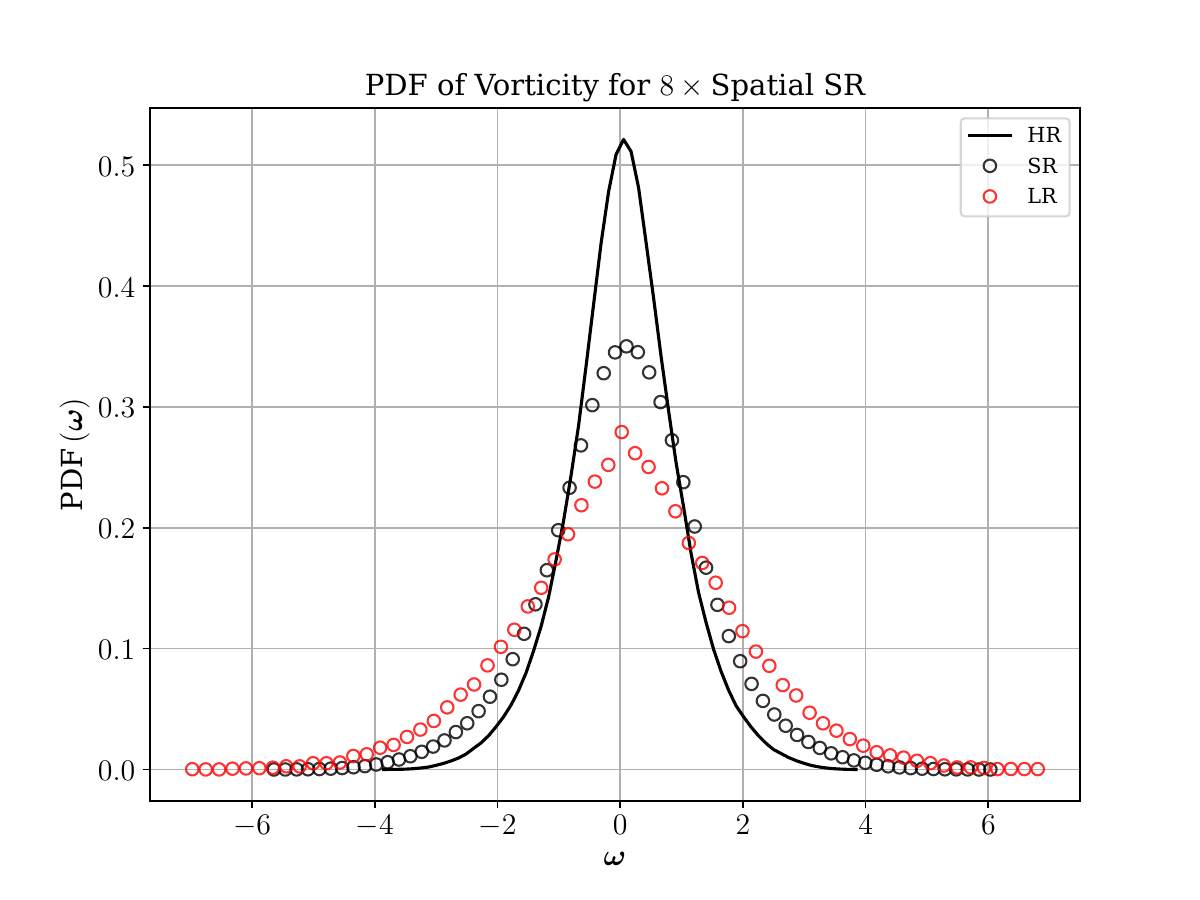}
  \setlength{\belowcaptionskip}{-3pt}
  \caption{Correlated-in-time Data}
  \label{fig: kol pdf bad}
\end{subfigure}
\centering
\begin{subfigure}{0.49\columnwidth}
\includegraphics[width=\textwidth, trim=5 0 5 0, clip]{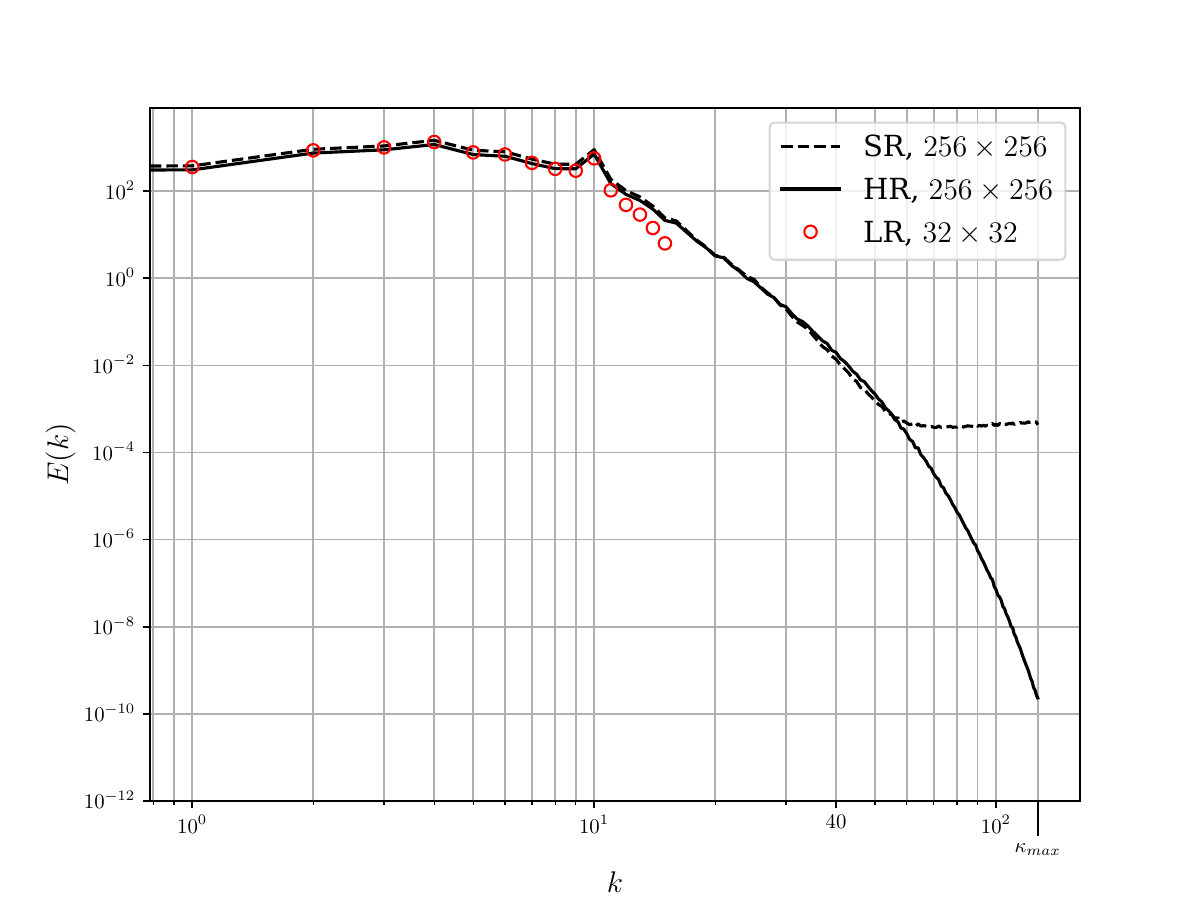}
  \caption{Decorrelated Data}
  \label{fig: kol spectra good}
\end{subfigure}
\begin{subfigure}{0.49\columnwidth}
\includegraphics[width=\textwidth, trim=5 0 5 0, clip]{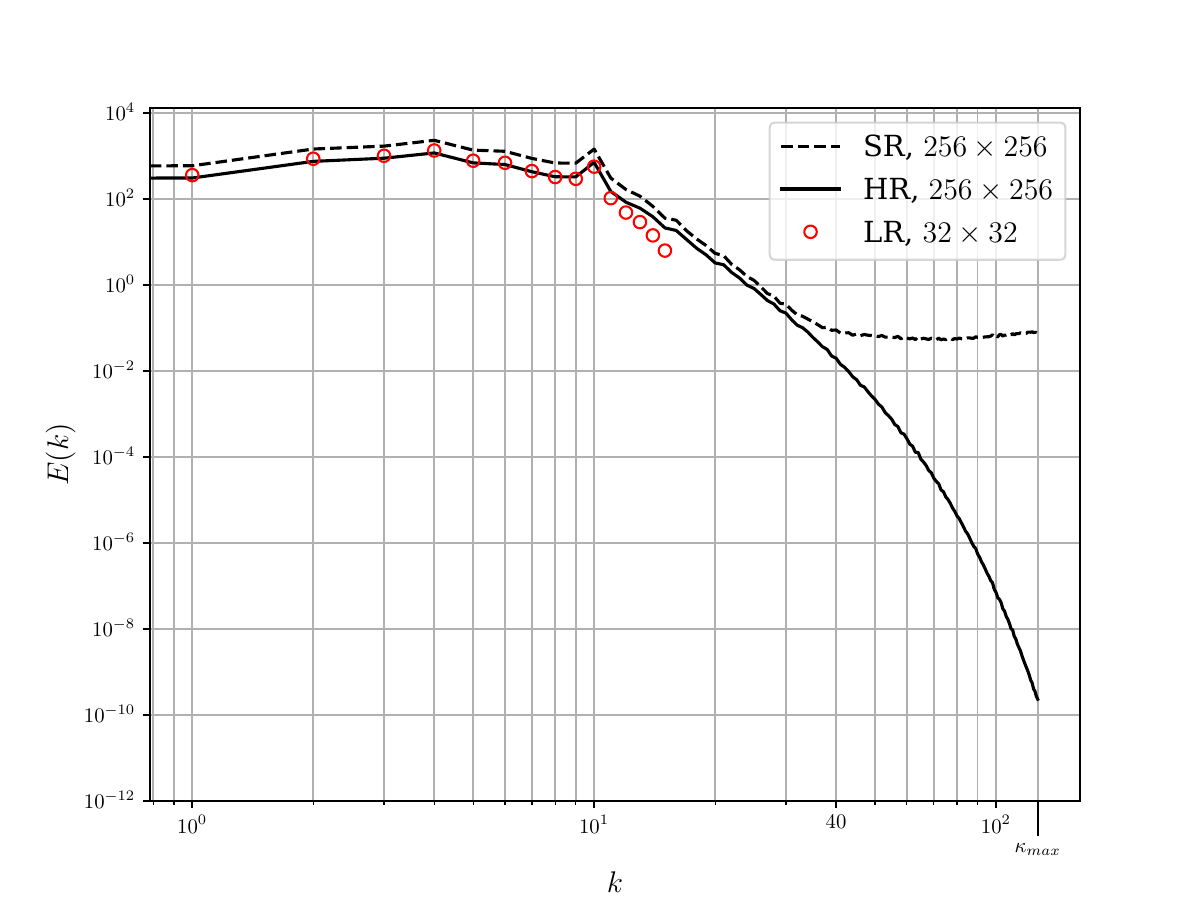}
  \caption{Correlated-in-time Data}
  \label{fig: kol spectra bad}
\end{subfigure}
\setlength{\belowcaptionskip}{-10pt}
\caption{$\textrm{PDF}\left(\omega\right)$ (top), and TKE Spectra (bottom), for the $32\times32$ LR, $256\times256$ SR, and $256\times256$ DNS.}
\end{figure}

Statistics of the generated snapshots can be used to quantify whether turbulent structures have been recovered well. The Probability Distribution Function (PDF) of the vorticity highlights the superior performance of the model trained on decorrelated data (Figure \ref{fig: kol pdf good}) in terms of capturing primary statistics of the underlying DNS data when compared to the model trained on correlated data (Figure \ref{fig: kol pdf bad}). Additionally, the time-averaged Turbulent Kinetic Energy (TKE) spectra of the results from both models clearly illustrate that the model trained on correlated-in-time data performs significantly worse than the model trained on decorrelated data (Figures \ref{fig: kol spectra good}, \ref{fig: kol spectra bad}) -- there is an observable offset in the spectrum throughout and tails off at lower wavenumbers. This example serves to highlight the impact that training on highly correlated data can have on model generalisability.

\section{On the Downsampling of DNS}
In this section we discuss the impact that downsampling of DNS data for tractability of training ML models can have on the perceived accuracy of those models. We illustrate this through the example of super-resolution on a Rayleigh-B\'enard convection cell. 

\subsection{Average-Pooling as an Effective Top-Hat Filter}


Training ML models on DNS datasets can be computationally intractable, as neural network based approaches require the tracking of gradients as data is passed through networks. This can be prohibitive as DNS datasets are often very large; even a smaller dataset, such as the JHTDB isotropic turbulence dataset, is still approximately 20 Tb.

Modern training of ML methods is typically carried out at scale on CUDA-enabled GPUs, the latest of which can possess more than 80 GB \citep{peng2024}. There are a number of methods widely employed in the machine learning community to facilitate computational tractability \citep{han2016b, howard2017a, zhang2017}, but the one we focus on is average-pooling \citep{lecun1989, lecun1998a}. 

\begin{figure}
\centering\captionsetup{width=\linewidth}
\includegraphics[width=0.4\textwidth]{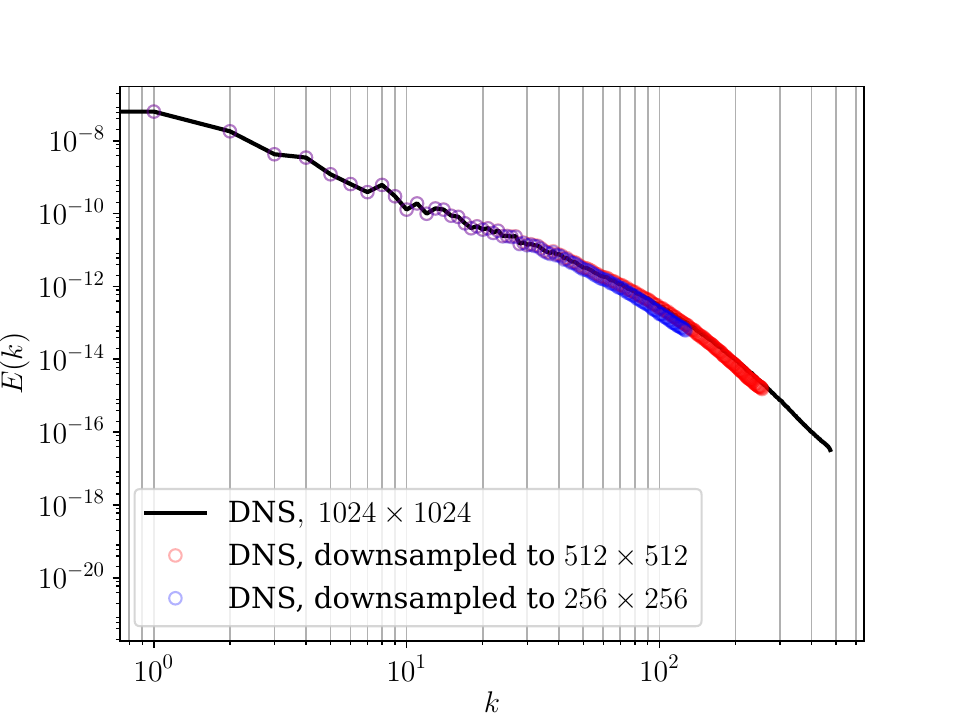}
\setlength{\belowcaptionskip}{-10pt}
\caption{TKE spectra for a given x-y plane of the JHTDB homogeneous isotropic turbulence (coarse) data.}
\label{fig: jhtdb spectra}
\end{figure}

Average-pooling allows for reduced data dimensionality, while filtering out small length-scale information (proportionally to the kernel width). Unlike natural images, where perceptual quality is the primary metric, turbulence can exist on length-scales much smaller than what the human eye perceives. We particularly draw focus to the effect that average-pooling can have on dissipation-range structures which define DNS as being fully-resolved, as in Figure \ref{fig: jhtdb spectra}, a time-averaged spectrum of the JHTDB coarse isotropic dataset. Average-pooling can also violate the divergence-free condition for incompressible flows, though spectral cut-off filters can be designed to preserve the divergence-free condition \citep{agdestein2025}.

\subsection{In prior work}




Two key approaches exist to address the issue of spectral cut-off due to average-pooling. The first is to work on problems where the grid size required for full resolution of the governing equations is small.
A common focus is 2D turbulence, such as 2D Taylor-Green vortices \citep{chen2021a}, decaying 2D homogeneous isotropic turbulence \citep{fukami2021b}, and 2D Kolmogorov flow \citep{kelshaw2022}. While this may be sufficient for the purposes of demonstrating methods, this approach effectively ignores issues of practicality in dealing with 3D DNS data for complex engineering flows. 
In other cases, 2D slices are taken from 3D fields 
\citep{liu2020a}
. In these instances, 2D CNNs are being used for 3D turbulence problems; we note that the impact on performance of passing 3D information through a 2D network remains to be explored. 


Several pieces of recent work address the computational constraints of training ML models on DNS by training on patches or subcrops of DNS fields, particularly for super-resolution \citep{bode2021a, jagodinski2023, yousif2022}. The advantages of doing so are that high-wavenumber turbulent structures are preserved and the effective dataset size is augmented, provided the two-point correlations do not indicate that the subsampled fields are highly self-correlated \citep{xu2023}. However, there is no guarantee that super-resolution models will produce samples which are coherent with one another, unless special measures are taken \citep{gao2015, xiang2023}.

\subsection{Results from a DDPM-based Super-Resolution of a Rayleigh-B\'enard Convection Cell}

Here we showcase our findings on the impact of average-pooling through a concrete example, using a DDPM to super-resolve snapshots of a 2D Rayleigh-B\'enard Convection case. The governing equations are the dimensionless incompressible Navier-Stokes equations in 2D, with the Boussinesq approximation for thermal-fluid coupling:

\begin{flalign}
&\dfrac{\partial u_i}{\partial t} + u_j \dfrac{\partial u_i}{\partial x_j} = - \dfrac{\partial p}{\partial x_i} + \sqrt{\dfrac{\text{Pr}}{\text{Ra}}} \dfrac{\partial^2 u_i}{\partial x_j \partial x_j} + T^* \delta_{i2} \hat{x}_2, && \\
&\dfrac{\partial T^*}{\partial t} + u_i \dfrac{\partial T^*}{\partial x_i} =\ \sqrt{\dfrac{1}{\text{RaPr}}} \dfrac{\partial^2 T^*}{\partial x_i x_i,} && \\
&\dfrac{\partial u_i}{\partial x_i} = 0,
\label{eq:RBC}
\end{flalign}

\noindent where $u_i$ is the velocity field, $x_i$ is the spatial coordinate, $p$ is the pressure, $T^*$ is a scalar temperature field, $\textrm{Pr}$ is the Prandtl number, $\textrm{Ra}$ is the Rayleigh number, $\hat{x}_2$ is a unit vector in the vertical direction, and $\delta$ is the Kronecker delta. 
The Rayleigh number is defined as $\textrm{Ra} = \frac{\beta \Delta T H^3 g}{\nu \alpha}$, where $H$ is the height of the cavity, $\beta$ is the thermal expansion coefficient, $\Delta T$ is the temperature difference between the top and bottom boundaries, $g$ is the gravitational acceleration, $\nu$ is the kinematic viscosity, and $\alpha$ is the thermal diffusivity. The Prandtl number is defined as $\textrm{Pr} = \frac{\nu}{\alpha}$. 
The domain is taken as a rectangle of aspect ratio 2, with unit height, $H$, discretized into $512\times1024$ spatial grid points, resolving to approximately $3 l_k$, where $l_k$ is the Kolmogorov length scale. Simulations are carried out using a spectral code \citep{burns2020}, with Fourier bases in streamwise directions and Chebyshev bases in non-periodic directions. The Prandtl number is set to $\textrm{Pr}=1$, and the Rayleigh number is set to $\textrm{Ra}=10 ^9$ to induce turbulent dynamics. We employ fixed temperatures at the top and bottom walls, and periodic boundaries in the streamwise direction. Generated data was decorrelated using the methods outlined in §\ref{subsec: corr prior work}.

\begin{figure}
\centering\captionsetup{width=\linewidth}%
\begin{subfigure}{0.44\textwidth}
    \centering\captionsetup{width=0.95\linewidth}%
    \includegraphics[width=\textwidth, trim=20 70 20 70, clip]{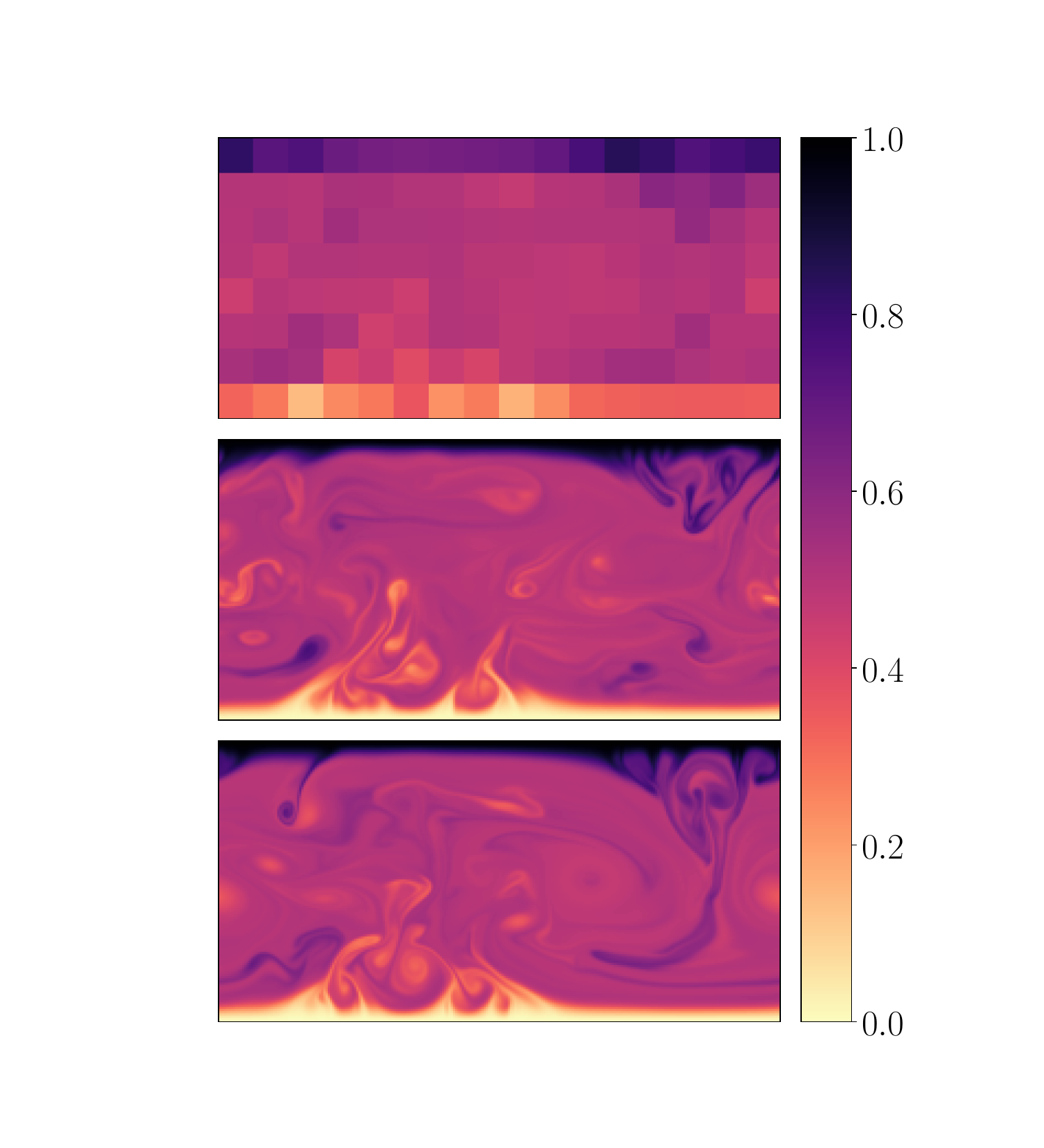}
    \subcaption{Temperature fields for $8\times 16$ low-resolution (top), $128\times256$ super-resolved (middle), $128\times256$ DNS (bottom).}
    \label{fig: too accurate rbc snapshots}
\end{subfigure}
\begin{subfigure}{0.44\textwidth}
    \centering\captionsetup{width=0.95\linewidth}%
    \includegraphics[width=\textwidth, trim=20 70 20 70, clip]{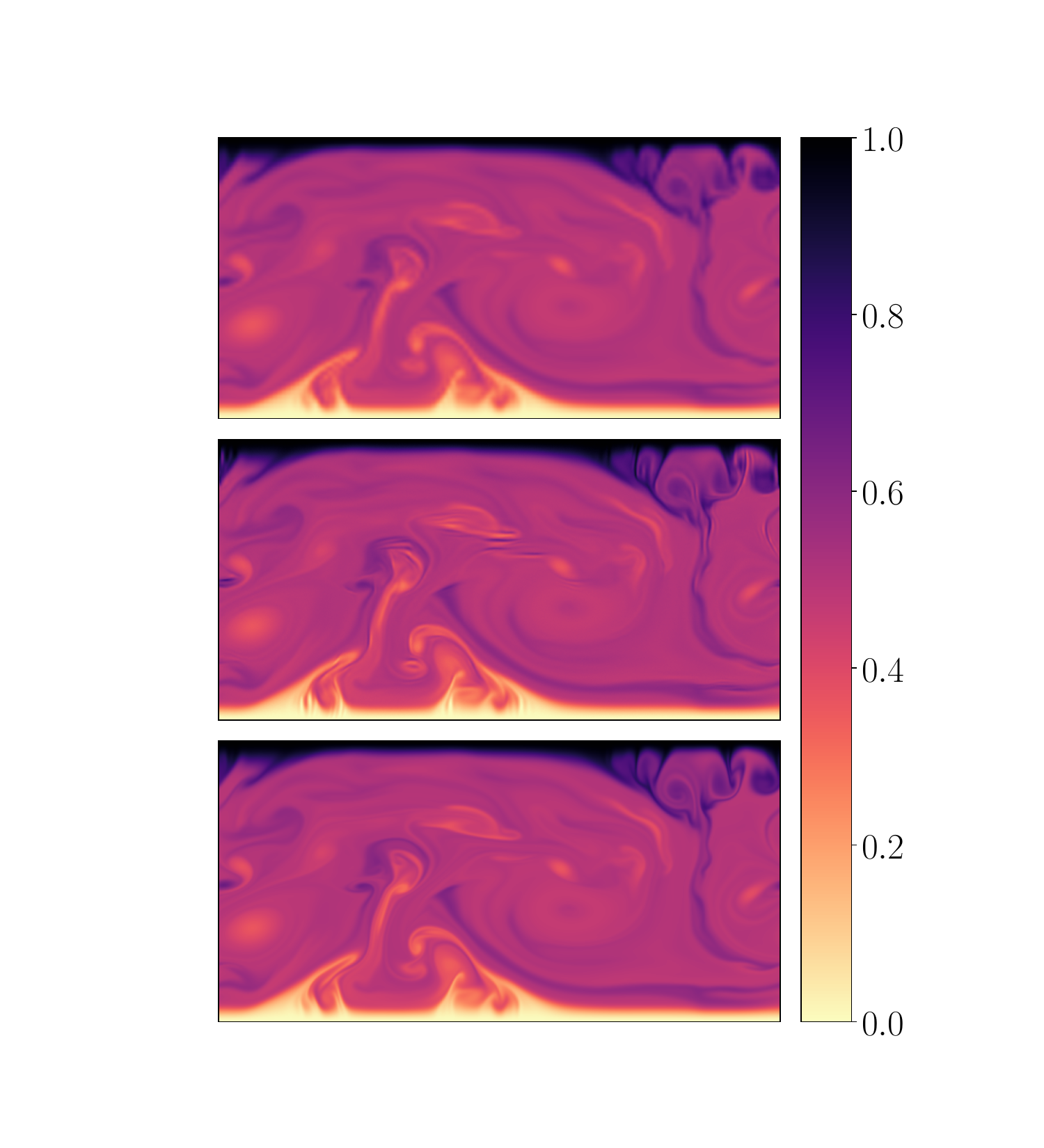}
    \subcaption{Temperature fields for $64\times 128$ low-resolution (top), $512\times1024$ super-resolved (middle), $512\times1024$ DNS (bottom).}
    \label{fig: less accurate rbc snapshots}
\end{subfigure}
\caption{Representative samples of super-resolved temperature fields from the pre-downsampled RBC case (left), and the full DNS RBC case (right). Note that the pre-downsampled RBC case is for $16\times$ SR, while the SR carried out on the full DNS is only $8\times$.}
\end{figure}

\begin{figure}
\centering\captionsetup{width=\linewidth}
\begin{subfigure}{0.44\textwidth}
    \includegraphics[width=\textwidth, trim=10 10 10 10, clip]{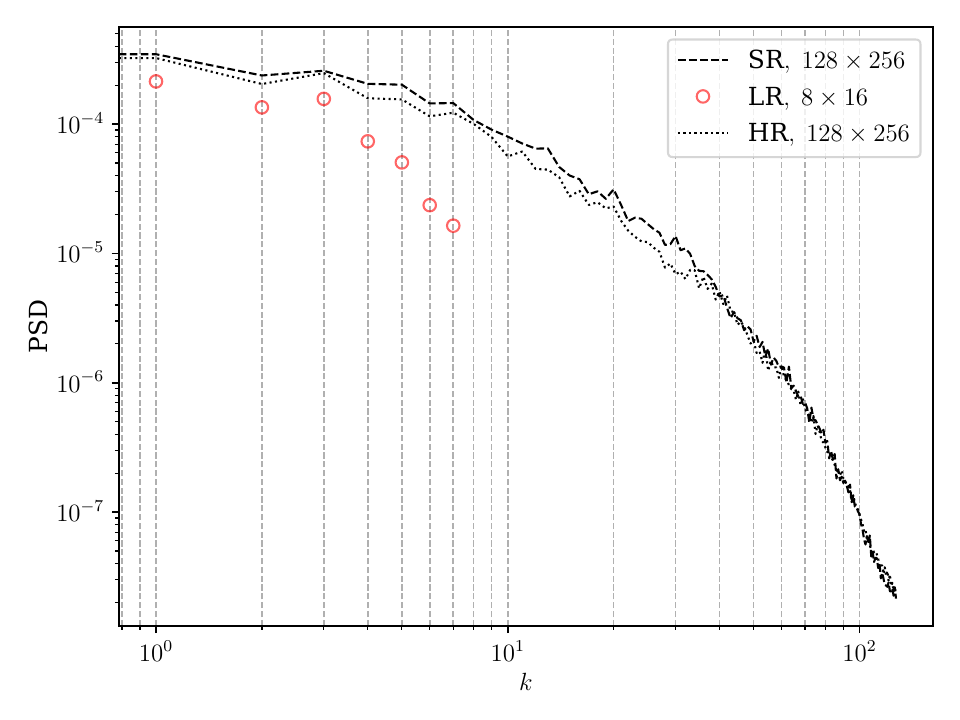}
    \subcaption{}
    \label{fig: too accurate rbc}
\end{subfigure}
\begin{subfigure}{0.44\textwidth}
    \includegraphics[width=\textwidth, trim=10 10 10 10, clip]{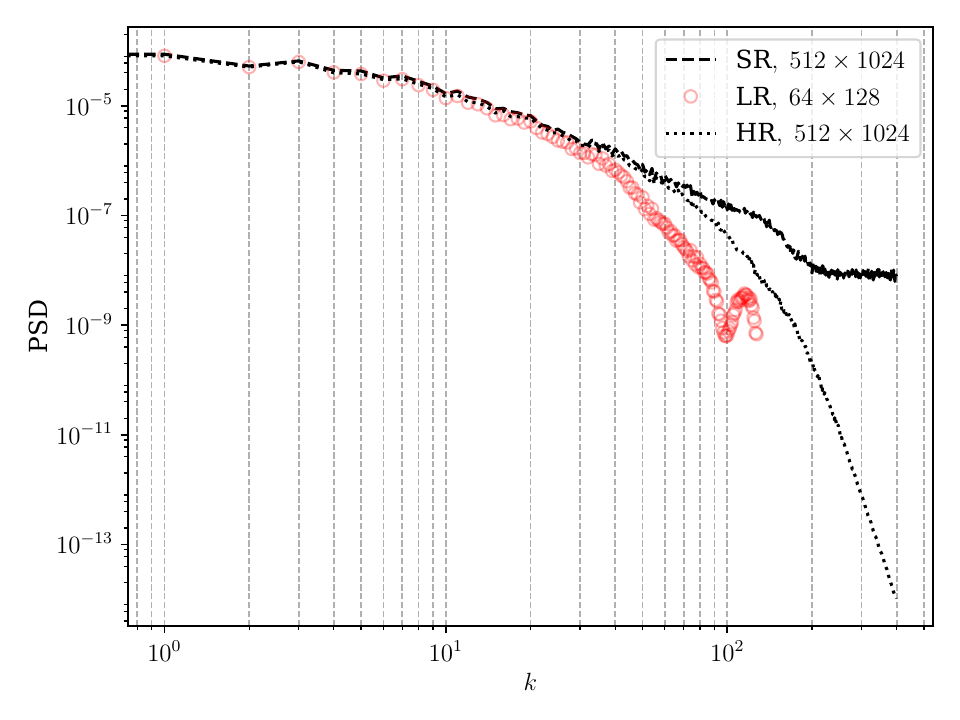}
    \subcaption{}
    \label{fig: less accurate rbc}
\end{subfigure}
\setlength{\belowcaptionskip}{-10pt}
\caption{Comparative PSDs of temperature fluctuations for the Rayleigh-B\'enard Convection case, at SR factors of: $16\times$ using DNS pre-downsampled from $512\times1024$ to $128\times256$ (left), and $8\times$ for the original DNS at $512\times1024$.}
\end{figure}

Initially, training a DDPM on DNS downsampled from $512\times1024$ grid points to $128\times256$ points led to seemingly very accurate model outputs. On an extreme super-resolution factor of $16\times$, the instantaneous snapshots of the reconstructed flow (Figure \ref{fig: too accurate rbc snapshots}) appeared to agree very well with the ground truth. The Power Spectral Density (PSD) in Figure \ref{fig: too accurate rbc} shows that from practically nothing, the PSD is recovered almost perfectly. Barring minor deviations from the true PSD, the high-wavenumber components of the flow are recovered well up to $k_{max}$, which was contrary to our expectations given the limited conditioning information.

\begin{figure}
\centering\captionsetup{width=\linewidth}
\includegraphics[width=0.4\textwidth, trim=10 10 10 10, clip]{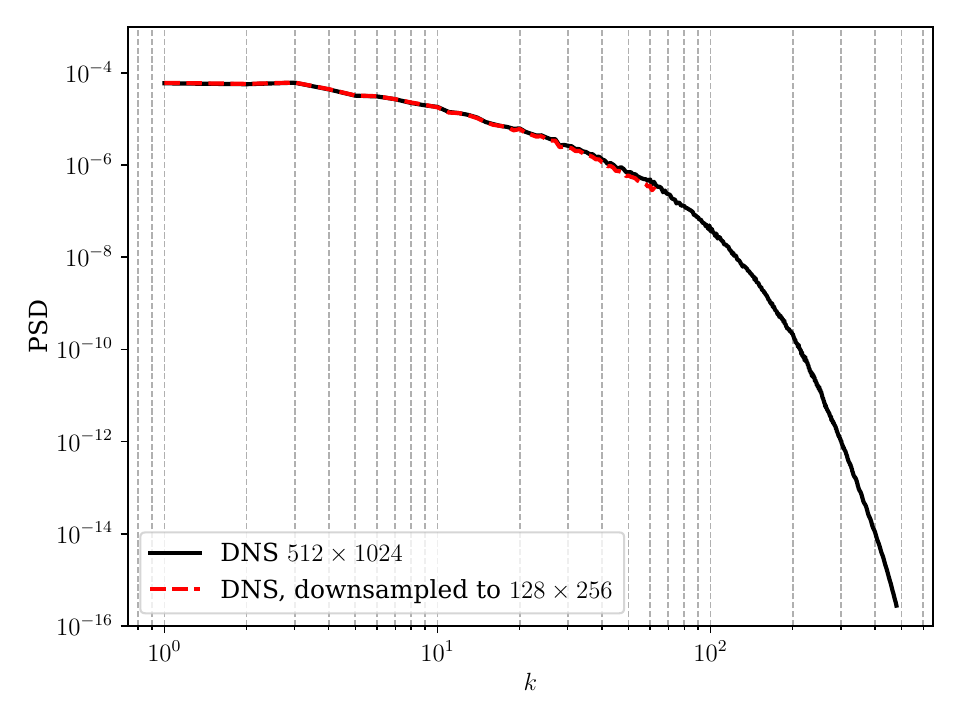}
\setlength{\belowcaptionskip}{-10pt}
\caption{An illustrative PSD of temperature fluctuations from the validation data used for the Rayleigh-B\'enard Convection case, with the truncating effect of average-pooling on the spectrum clearly visible.}
\label{fig: effect of downsampling}
\end{figure}

On spectral analysis of the original DNS against the downsampled DNS, it was observed that the dissipation range was being cut off by downsampling (Figure \ref{fig: effect of downsampling}), trivialising the super-resolution problem. Further machine learning experiments, carried out at the original DNS resolution of $512\times1024$, show a clearer picture. From Figure \ref{fig: less accurate rbc snapshots}, it is easier to observe that although the flow has been reconstructed well, there are deviations from the DNS. The PSDs in Figure \ref{fig: less accurate rbc} show that while the DDPM reconstructed fields (SR) are accurate up to $~k=60$, the DNS extends to wavenumbers much past this point -- and shows a clear dissipation range. We emphasise here that drawing conclusions from Figure \ref{fig: too accurate rbc} may lead to misleading claims of state-of-the-art performance, when the performance on full resolution DNS is more modest. 

We note that there is a slight discrepancy in performance up to wavenumbers of $k=100$ from the $16\times$ SR (Figure \ref{fig: too accurate rbc}) to the $8\times$ SR (Figure \ref{fig: less accurate rbc}). This is attributed to a fixed
model architecture used for both resolutions. Proportionally, the $8\times$ SR task will ideally have been carried out using a larger DDPM-parameterising network, but the points of comparison are still justified. 


\section{Conclusions and Further Work}

In this work, we have provided an overview of two points of concern in the use of DNS turbulent flow data for ML. The presence of correlations-in-time in turbulence, leading to persistent structures, can lead to highly similar snapshots of data being presented to networks during training as unique samples. While this adds redundancy to training datasets, correlations across training and validation can actively hinder the training process. This was demonstrated on a super resolution task for a Kolmogorov flow using a DDPM, where it was made clear that training on correlated-in-time data significantly decreased the performance of a model. However, developments in ML have allowed for the incorporation of temporal dependencies in some instances \citep{xiong2019}, and spatiotemporal models are actively being investigated for turbulence applications \citep{fukami2021b}. 

It was observed that average-pooling/downsampling of data -- a common dimensionality reduction technique for computational tractability of ML training -- can cause problems with DNS datasets. The average-pooling operator is effectively a spectral cut-off filter. It was shown that for certain DNS cases, downsampling led to the dissipation range of wavenumbers being filtered out, the presence of which is typically what defines DNS as being DNS (in that all scales of turbulence are resolved). The impact this may have on ML models was investigated using the example of DDPM-based super-resolution of a Rayleigh-B\'endard Convection cell. It was found that downsampling the DNS made the model appear to perform very well, but this was merely an obfuscation of the true model performance. Re-training the model at the true DNS resolution led to more realistic results, which would be more meaningful for any intended application of the ML methods.

\backsection[Funding Statement]{This work was supported by the Engineering and Physical Sciences Research Council [EP-T517823-1]. SD acknowledges a Dame Kathleen Ollerenshaw Fellowship.}


\bibliographystyle{jfm}
\bibliography{references.bib}

\end{document}